\documentclass[useAMS,usenatbib]{mn2e}
\usepackage{graphicx}
\usepackage{amssymb}
\usepackage{times}
\usepackage{subfigure}
\usepackage[T1]{fontenc}
\usepackage{aecompl}

\defcitealias{Nissen2010}{N10}

\newcommand{\fig}[1]{Fig.~\ref{#1}}
\newcommand{\sect}[1]{Sect.~\ref{#1}}

\begin{document}

\title[Distances of twin stars]{Climbing the cosmic ladder with stellar twins}
 \author[P. Jofr\'e et. al.]{P. Jofr\'e$^{1}$\thanks{E-mail: pjofre@ast.cam.ac.uk}, T. M\"adler $^{1}$, G. Gilmore$^{1}$, A. Casey$^{1}$, C. Soubiran$^{2}$ and C. Worley$^{1}$  \\
$^{1}$Institute of Astronomy, University of Cambridge, Madingley Road, Cambridge CB3 0HA, UK\\
$^{2}$Univ. Bordeaux, CNRS, Laboratoire d'Astrophysique de Bordeaux, F-33270, Floirac, France}

\date{Accepted .... Received ...; in original form ...}


\maketitle

\label{firstpage}

\begin{abstract}
Distances to stars are key to revealing a three-dimensional view of the Milky Way, yet their determination is a major challenge in astronomy.    Whilst the brightest nearby stars benefit from direct parallax measurements, fainter stars are subject of indirect determinations with uncertainties exceeding 30\%. We present an alternative approach to measuring distances using spectroscopically-identified twin stars. Given a star with known parallax, the distance to its twin is assumed to be directly related to the difference in their  apparent magnitudes. We found 175 twin pairs from the ESO public HARPS archives and report excellent agreement with Hipparcos parallaxes within 7.5\%.   Most importantly, the accuracy of our results does not degrade with increasing stellar distance. With the ongoing collection of high-resolution stellar spectra, our method is well-suited to complement Gaia.

\end{abstract}

\section{Introduction}
\label{sec:Introduction}

Knowing the distances of stars is  fundamental for many subjects in astrophysics.
Indeed,  distances are required to understand the evolutionary status of stars and place them onto the   Hertzprung-Russell (HR) diagram  \citep[e.g.][]{1967Sci...155..785I, 1967ApJ...147..624I, 2014EAS....65...17C}, which relates their effective temperature with their intrinsic luminosity. 
Although the effective temperature is easily determined by several observational techniques from spectroscopy \citep[see e.g.][for a review]{1992oasp.book.....G} or from photometry \citep[e.g.][]{1999A&AS..140..261A},  the intrinsic luminosity is related to the flux emitted by the star, which is inescapably coupled with the distance. 
Precisely knowing the distance and temperature of a star allows us to separate its space motions, estimate physical properties such as radius, mass, age, and to infer internal nuclear reaction yields. Combining this information for many stars allows us to understand the structure and evolution of the Milky Way.

The HR diagram has been successfully used for cluster stars \citep[e.g.][]{2003A&A...408..529G, 2014EAS....65...17C} because these stars are believed to have the same age, chemical composition and distance, and relations between their magnitudes and luminosities can be employed.   For individual stars in the Galactic field, the HR diagram can be utilised with confidence only for stars for which a parallax has been measured \citep{2008A&A...488..935G, 2014EAS....65...17C}.  These stars correspond to the closest and brightest ones in the solar vicinity observed by the Hipparcos mission \citep{1997A&A...323L..49P}.  Very soon the Gaia mission  \citep{2001A&A...369..339P} will provide parallaxes and proper motions with unprecedented accuracy for one billion stars in the Galaxy, allowing us to construct HR diagrams beyond the solar vicinity.

The fields of  stellar and Galactic astrophysics are converging into one field thanks to the  Gaia mission 
and its complementary ground-based high-resolution spectroscopic surveys, such as Gaia-ESO \citep{2012Msngr.147...25G}, 4MOST \citep{2011Msngr.145...14J}, RAVE \citep{2006AJ....132.1645S}, APOGEE \citep{2008AN....329.1018A} or GALAH \citep{2015arXiv150204767D}.  
In this pre-Gaia era, methods to determine the distance of typical Galactic field stars beyond Hipparcos are constantly under development \citep{2008ApJ...684..287I, 2010MNRAS.407..339B, 2014MNRAS.437..351B,2014MNRAS.445.2758R, 2014ApJ...784..170X}. Most methods, commonly also referred to as ``spectroscopic parallaxes",  are indirect as they rely on parameters determined from colours or from spectra, and of course on stellar evolution models, to predict the position of stars in the HR diagram.  The estimated intrinsic luminosity of a star is used to estimate the distance, given its apparent brightness on the sky. The power of these methods is that they provide  distances estimates for large numbers  of field stars. Nonetheless, they require assumptions { about} extinction along the line-of-sight, but more seriously they are limited by systematic uncertainties in stellar evolution modelling and spectroscopy, which can result in distance  errors of the order of up to 70\% for metal-poor giants \citep{2013MNRAS.429.3645S} { and many other spectral types}.

Other attempts to determine distances with better precision use particular types of stars with well defined intrinsic luminosities such as variable stars, blue horizontal branch stars or red clump stars \citep{2008A&A...481..441G, 2014ApJ...796...38N}, but the luminosity of those stars is  also subject to calibrations from stellar evolution models.  An alternative  idea is to compare the spectra of stars with those for which  the parallax is known.  \cite{2003A&A...398..141S} introduced this concept to determine the atmospheric parameters and absolute magnitude of a star by computing a likelihood function of the difference between its spectrum and a grid of observed spectra of stars with parallaxes. The reported distances errors are less than 30\% for when compared with Hipparcos parallaxes of FGK stars \citep{2011A&A...525A..90K} and less than 12\% when red clump stars are considered \citep{2008A&A...480...91S}. This compares well with other methods but in this case the distances rely on interpolations in the parameter space to compute the likelihood function and not on stellar evolution and atmosphere models.

We report another perspective of the concept of direct comparison of spectra used in \cite{2003A&A...398..141S}, but here we focus only on twin stars. We call this method  ``the twin method" and distances (parallaxes) determined with it are referred to as ``twin distances" (twin parallaxes).  The method is based on two principles: 

\begin{itemize}
\item Twin stars are { proposed} to have the same luminosity. We identify such stars by comparing their spectra, which must be identical {according to a criterion described in \sect{twins}.}.
\item Because of our proposition, the difference of the apparent magnitudes of twin stars { is directly related to} the difference in their distances. { Thus,} knowing the distance of the first star from its parallax, allows us to { directly ascend on the cosmic ladder}.  
\end{itemize}

{ For details on the mathematical formalism of this method we refer the reader to the Appendix~\ref{formalism}.  The twin method is different from other methods to determine stellar distances}, in particular from the regular spectroscopic parallax method, because its { foundations}:  First, we do not require a spectral parameter determination method to analyse a star spectrum with the aim to obtain its effective temperature, surface gravity and metallicity. Spectroscopic parallax methods need this information to place the star in the HR diagram which, with the help to an evolutionary model, a luminosity can be assigned to the star.  In fact, we do not use isochrones or the HR diagram to determine the distances. Second,   
in our method we do not determine the value of the absolute magnitude (or distance modulus) of individual stars, instead we use the relative magnitudes to work out the distance directly.  Third, when using twin stars, the difference of their de-reddening is directly related to the observed colours of the twins, which is useful because we do not need to use extinction maps to correct { reddened} magnitudes. Furthermore, we consider several photometric bands in our distance calculation, allowing us to treat the effects of dust extinction { more robustly as well as to assess cases with bad photometry in a given band}.

Although very soon Gaia will provide accurate parallaxes for the majority of the stars observed with high-resolution by current instruments, we fully agree with the point made by  \cite{2010MNRAS.407..339B}, that other distance estimation methods complementing these parallaxes are crucial, especially for more distant stars. Moreover, it will remain important to have alternative distance measurements of stars that are independent from parallaxes or from model-dependent methods. This is  especially true for stars located in remote open clusters, for which parallax measurements will remain challenging even for Gaia, making isochrone fitting to their colour-magnitude diagrams one of the only  available methods to determine their distances.   

In this article we introduce the twin method and analyse its performance in a sample of FGK  stars with Hipparcos parallaxes { as well as in open clusters which have} spectra taken with high resolution. Applications to lower resolution spectra and to stars beyond the reach of Hipparcos will be presented in future works on this subject. 



\begin{figure*}
\begin{center}
\includegraphics[scale=0.6]{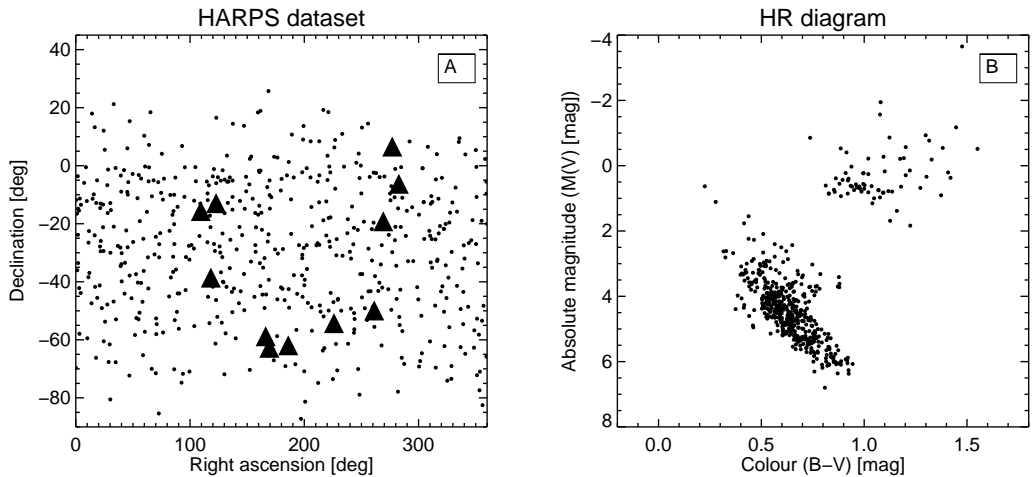}
\caption{Our sample of Hipparcos stars with HARPS spectra.  A: Distribution of stars across the observable Southern sky from La Silla Observatory, Chile. Triangles correspond to positions of clusters. The sinusoidal shape drawn by the open clusters reflects the location of the Galactic disk. B: HR diagram of the Hipparcos sample. The absolute magnitudes were computed using the provided parallax  and the apparent magnitudes $B$ and $V$ are in the Johnson-Cousins filter system.  The main-sequence and red giant branch are visible. }
\label{fig:hr}
\end{center}
\end{figure*}

\section{Spectroscopic and photometric sample}\label{data}

{ To test the twin method we require stars to comply the following criteria: 
\begin{itemize}
\item Parallaxes from Hipparcos with accuracies of better than 10\%. 
\item Stellar parameters of typical FGK Milky Way stars. For that we took the parameters from the literature as compiled {in} the PASTEL catalogue  \citep{2010A&A...515A.111S}. For main-sequence stars we selected from PASTEL all stars that have temperatures in the range [5000, 7000]~K and surface gravities in the range [3.5, 5.0]~dex. For giants we selected from the PASTEL database stars in temperature range  [3500,5000]~K and surface gravity range [0.5, 2.0]~dex. 
\item They are not classified as variables by the SIMBAD database \citep{2000A&AS..143....9W}.
\item The stars are not SB2 binaries, i.e. no double spectroscopic lines in the analyses reported by the literature compilation of PASTEL. 
\item They have photometry {in five photometric} bands.
\item They have been observed with the HARPS spectrograph. 
\end{itemize}
}
This gives us a total of 536 field stars, whose positions  in the sky and in the HR diagram can be seen in  Fig.~\ref{fig:hr}. The left panel shows their distribution on the sky, represented with dots. The triangles correspond to the open clusters initially selected for our study (see below). The stars are well distributed in the Southern Hemisphere.   The right panel shows the HR diagram of our sample. For this illustration, the absolute magnitude was calculated using the parallaxes of \cite{2007A&A...474..653V} and the $V$ magnitudes and $B-V$ colours are corrected by dust extinction using the values provided in \cite{1997ESASP1200.....E}.  The gap between the main-sequence and the red-giant branch is a purely selection bias since we selected main-sequence and giants separately using temperature and logg cuts { (see above)}.   

\subsection{Spectra}


The HARPS instrument is fibre-fed by the Cassegrain focus of the 3.6m telescope in La Silla \cite{2003Msngr.114...20M}. The spectra were reduced by the HARPS Data Reduction Software (version 3.1) and were downloaded from the advanced data product from the ESO archives.  We chose this instrument because of the  very high signal-to-noise ratio (SNR), high resolution and long wavelength coverage that a typical bright Hipparcos star spectrum has when observed with HARPS.

The fact that the majority of the stars in our sample are on the main-sequence (see HR diagram of Fig.~\ref{fig:hr}) is an observational bias, due to the many projects looking for planets in solar analog spectra in the HARPS database. Our sample includes the solar spectrum  from the Ceres asteroid which has a signal-to-noise (SNR) of $\sim 250$ for calibration purposes of our method to search for twins, { i.e. we used it to see if we could find the solar twins in our sample that have been studied in other works. } This spectrum was taken from the library of Gaia FGK benchmark stars\footnote{http://www.blancocuaresma.com/s/benchmarkstars/} \citep[see][for a summary]{2014ASInC..11..159J}, which was processed in the same way as the rest of the data of this study (see \sect{s:ispec}). Our final sample contains 536 FGK { stars}.

Finally, we collected spectra from stars belonging to 11 open clusters analysed by \cite{2015A&A...577A..47B}.  { The selection criteria for cluster stars was different than for the field stars, in the sense that no {colour} cut or another selection on the atmospheric parameters  was done}. We queried all spectra  for cluster stars in the HARPS archives, with the only restriction to have  SNR of at least 10. For cluster stars we stacked the spectra of the same star to increase the SNR. Our final sample includes 177 open cluster stars.

\subsection{Photometry}\label{photometry}

To determine the distance from the apparent magnitudes of stars it is important to have as many photometric bands as possible. This is crucial in our analysis, as the distance is obtained from the difference of magnitudes, which needs to yield consistent results when different photometric bands are taken into account.  Thus, when the results obtained with different photometric bands agree, we ensure the reliability of the distance obtained, as well as the photometric quality of the star.  Another very important aspect of having several photometric bands for our analysis is the { effect} of extinction { or saturation}, which is different for the different bands.  When determining distances from different photometric bands, the effects of extinction { or saturation} are less dependent on our final results. { Furthermore, it is possible  that some stars are missclassified variable stars. {Different photometric measurements are useful for diagnosing if this
is the case.}}  For these reasons, the determination of distance was done using the  photometric bands from Hipparcos {\it Hp} \citep{2007A&A...474..653V}, Johnson-Cousin {\it B, V} \citep{1997ESASP1200.....E}  and  2MASS {\it J, Ks} \citep{2006AJ....131.1163S}. 

The ratio of total-to-selective extinction $\mathrm{R}(X)$ was assumed to have five different values corresponding to the five different bands as taken from \cite{1999PASP..111...63F} and \cite{2013MNRAS.430.2188Y}.  The values are  $\mathrm{R}(H_p) = 3.1$, $\mathrm{R}(B) = 3.8$, $\mathrm{R}(V) = 3.1$,  $\mathrm{R}(J) = 0.7$ and $\mathrm{R}(Ks) =0.3$.    Note that the reddest $Ks$ band has a very small extinction, with $\mathrm{R}(Ks)$ of 0.3, while the bluest $B$ band has a significantly larger extinction, with $\mathrm{R}(B)$ of 3.8. 
{ It is important to remark that our} present analysis { for field stars} is restricted to Hipparcos stars, which are close by and are not significantly affected by extinction \citep{2003A&A...411..447L}.  The stars of our sample are mostly nearby disk stars, and the assumption of constant $\mathrm{R}$ seem to be reasonable in the five photometric bands considered. We will see in \sect{clusters} that using five bands also help us to treat extinction properly in open clusters. { Hence, we prefer to use our generic form consistently  throughout our work by applying the extinction correction factor of Eq.~\ref{plx} in our analysis. }

\section{Procedure to find spectroscopic twins}\label{twins}

\subsection{Spectral pre-processing}
\label{s:ispec}
The spectra needed to be pre-processed before the analysis. We carefully normalised the spectra, corrected for the radial velocities of each star,  removed telluric absorption, and resampled the spectra to a common wavelength range using the tools described in \cite{2014A&A...566A..98B}.  {This was done
homogeneously in order to ensure that the relative comparison of the spectra is as objective as possible.}


The normalisation was done by fitting splines of second order locally to the spectra as described in  \cite{2014A&A...566A..98B}.  The radial velocity correction was done by cross-correlating the spectra against a template rest-frame. This template was either an atlas of the spectrum of the Sun or of Arcturus, depending on the colour of the star.  If the star had $B-V < 1.0$~mag, the star is most probably a dwarf and we used the template of the Sun. Otherwise if the star had $B-V > 1.0$~mag, then the star is probably a giant and the template of Arcturus was more suitable for the cross-correlation.  Both atlases are included with the tools to prepare the library of \cite{2014A&A...566A..98B}.

 \begin{figure*}
\begin{center}
\includegraphics[scale=0.5]{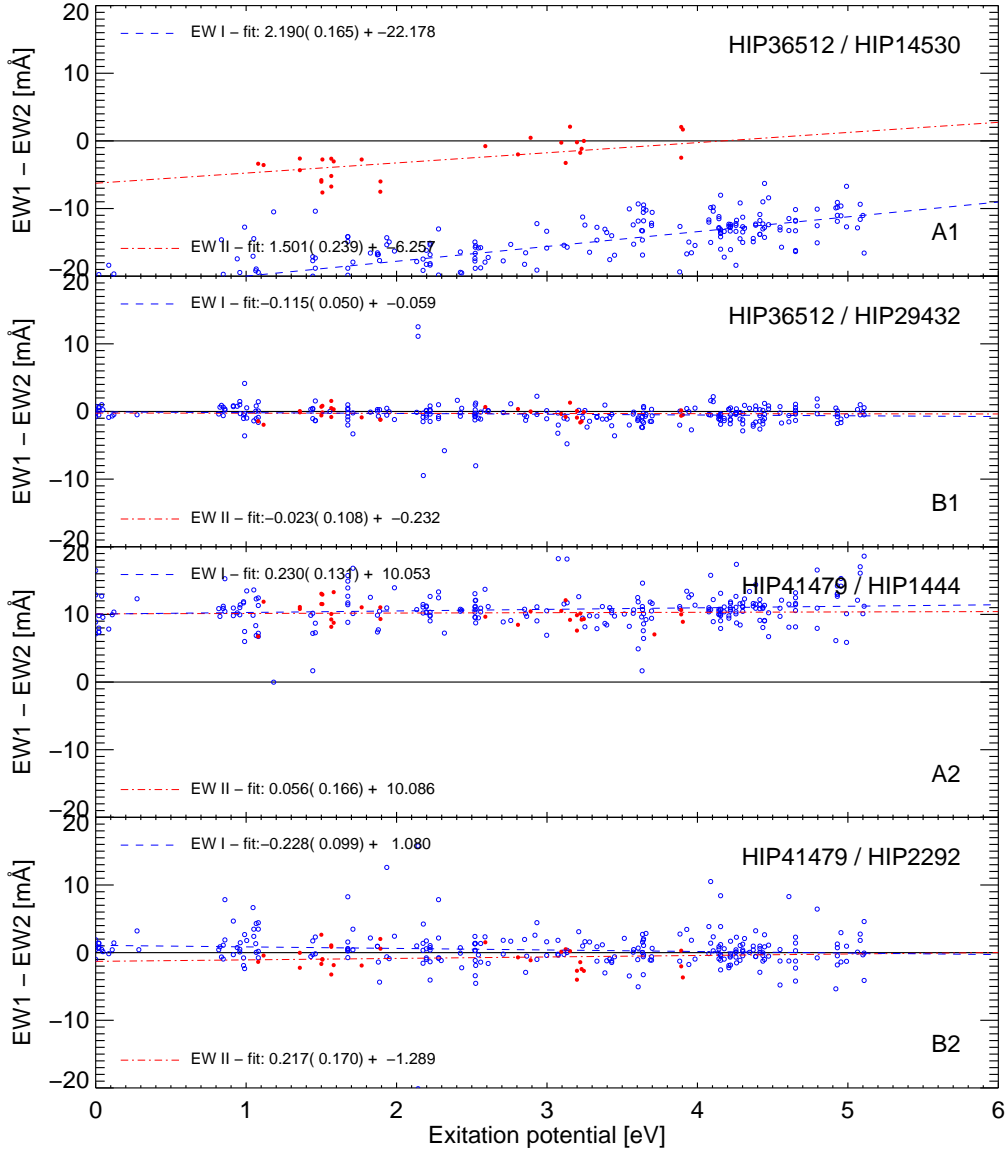}
\vspace{-2.0cm}
\caption{Example of equivalent width differences of four  {pair of stars} as a function of excitation potential. A linear regression fit is performed to the data of neutral lines (blue open circles) and ionised lines (red filled circles), separately. The slope and zero point of the fit is indicated in each case, with the { error} of the slope in parenthesis. The {names of the stars from the pair} are indicated in the right hand side of each panel. {Two examples are shown, where panels A1 and A2 show regular stars and panels B1 and B2 show twin stars}.}
\label{fig:ilis}
\end{center}
\end{figure*}

\subsection{Spectral analysis}\label{s:ilis}

{ Following previous spectroscopic studies {of} solar twins \citep[e.g.][]{2006ApJ...641L.133M, 2014A&A...572A..48R, 2014MNRAS.439.1028D} we measured the strength of atomic absorption lines, i.e. we worked with equivalent {widths} (EWs) by differentiating each  and studied them as a function of excitation potential (E.P.). 
To account for broadening effects caused by e.g. rotation,  we also investigated the depth of each line by calculating the value of their  minima with respect to fluxes at $\pm 0.02$~nm from the centre of the line, similar to what \citet[e.g.][]{2006ApJ...641L.133M} have done.

  We employed the atomic data for Fe, Si, Mg, Ca, Ti, Sc, V, Cr, Mn, Co and Ni used for the Gaia benchmark stars \citep{2014A&A...564A.133J, 2015arXiv150700027J}, which provide a substantial number of transitions in FGK stars. We calculated EWs of 423 atomic lines listed in these papers, which have been carefully selected
to be unblended lines for typical FGK Milky Way stars.  To assess the systematic error of the EW measurement, we employed two different { codes to measure the EWs}:  {\tt iSpec} \citep{2014A&A...569A.111B} and {\tt SMH} \citep{2014arXiv1405.5968C}.  We selected the lines with EWs between 15 and 120 m\AA\  showing a good agreement between both methods. These lines are strong enough to have accurate EW measurements in both methods, but weak enough for the line not to saturate.   The EWs were then compared between two spectra, which reduced our data size from 262,000 pixels to $<$423 points (per star), facilitating a faster and focussed comparison.

 We analysed the trend and offset of the differences of EWs as a function of E.P. for each combination of spectra, which was done by fitting a line to the data by linear regression.  To obtain a robust  estimate of slope of our linear fit, we performed bootstrapping by fitting a dataset of  randomly selected 80\% of lines 1000 times. The final slope was the mean of the 1000 fits obtained, with its standard deviation as its uncertainty.   

\begin{figure}
\hspace{-1cm}
\includegraphics[scale=0.45]{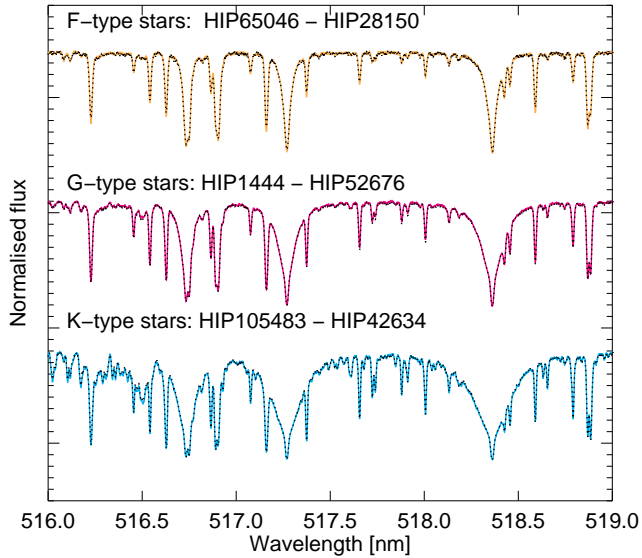}
\caption{Example of twins observed with HARPS for stars with FGK spectral classes. Coloured continuous lines represent one star, and black dotted lines represent the twin. Effective temperature decreases from top to bottom. }
\label{fig:twin}

\end{figure}
Twin stars are required to have fits with zero slope and zero offset {(value of the fit at E.P = 0)} of both, ionised and neutral lines. Since the linear fits have  a dispersion around zero, we carefully studied the values of slopes and offsets   for our complete dataset. After trying several different combinations and performing visual inspection of the selected twin pair spectra, we found that a good compromise of spectroscopic twins and number of selected pairs was obtained under the following criteria: 
}
\begin{enumerate}
\item {The slope and offset do not differ by more than }0.3 m\AA/eV  and  2.5 m\AA\ for neutral lines, respectively. 
\item Error of the slope of neutral lines of less than 0.1 m\AA/eV.
\item {The offset between the ionised and neutral lines is not more than} 2.5 m\AA\ for ionised lines with respect to neutral lines.
\item Mean difference of line depths of less than 10\%.  
\end{enumerate}


Examples of how EWs compare for { different}  stars are found in Fig.~\ref{fig:ilis}.  The open blue circles represent neutral lines and the filled red circles represent ionised lines. The regression linear fits to the data are  represented with blue and red dashed lines, for the neutral and ionised lines, respectively.  The values of the slope and offset are indicated, with the { error} of the slope  in parentheses. {   The first two panels show the star HIP~36512 compared to a regular star {(Panel A1)} and a twin star {(Panel B1), respectively. The last two panels  show another example, but for the star HIP~41479, {in which Panel A2 and B2 show a regular and a twin star, respectively}. For the regular cases {(Panels A1 and A2)},  we see clear trends and offsets in the linear fits, while for the twin stars {(Panels B1 and B2)}, the fits follow the zero line,  as defined with our criteria above.  }

Three examples of twin pairs identified using our method can be seen in Fig.~\ref{fig:twin}. Each example shows a spectral region around the Mg I b triplet, a feature known to be very sensitive to stellar parameters.  For this illustration, one star of each twin pair is plotted with a continuous coloured line, and the other one is plotted with black dotted line. The spectra are { indistinguishable within the errors}; without knowing the coordinates and magnitudes, it is impossible to disentangle if both spectra correspond to two independent stars or the same one observed in different conditions.

A total of { 175} twin pairs were found using these criteria. These pairs include triplets or quadruplets, partly because of the biases in the HARPS sample towards solar analogs.  Although this number seems very large, it is small given the total amount of pairs analysed in this work ($n \times (n+1) /2$ where $n$ is the number of stars in the dataset.) {In this particular sample, we find 175 twin pairs out of a total of 12844 comparisons, i.e.  1.36\% of the total pairs resulted in twin stars.  } Their location in the HR diagram can be seen in Fig.~\ref{fig:hr_twins}.  The filled circles (blue and red) are stars with twins in our sample that satisfy the spectroscopic criteria. The red circles correspond to our final set of twins, which had good photometry  (see \sect{photometry}).  From \fig{fig:hr_twins} we can see that most of the twins found are well distributed in the main-sequence, although some of them are subgiants and red clump stars. We could not find twins for the cool giants of our sample, which is not surprising when seeing how they distribute in the HR diagram. One can find few stars with similar colour and luminosity at these temperatures.  {The few
spectroscopic twins found on the red giant branch} had photometry in different bands that was not consistent enough for reliable distance measurements.  
 \begin{figure}
\hspace{-1cm}
\includegraphics[scale=0.45]{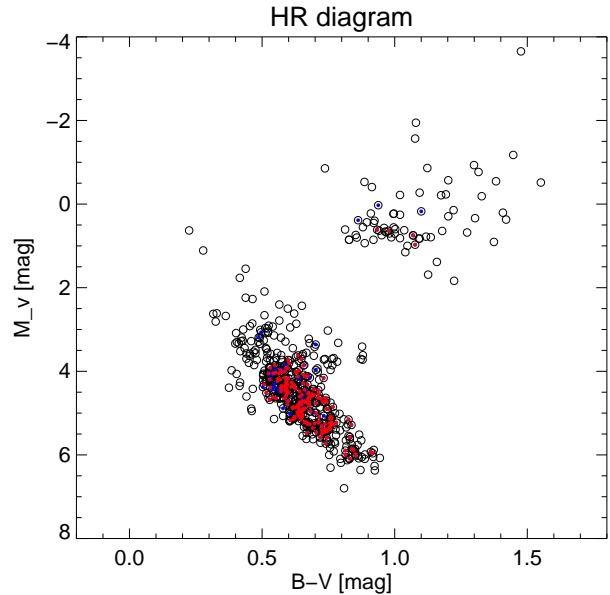}
\caption{HR diagram of the Hipparcos sample. Open circles: the entire sample of 536 stars, filled circles: stars with twins. Blue colour corresponds to spectroscopic twins while red colour indicates the stars for which a distance could be determined with a good consistency over several photometric bands. }
\label{fig:hr_twins}
\end{figure}

 \begin{figure*}
\begin{center}
\includegraphics[scale=0.5]{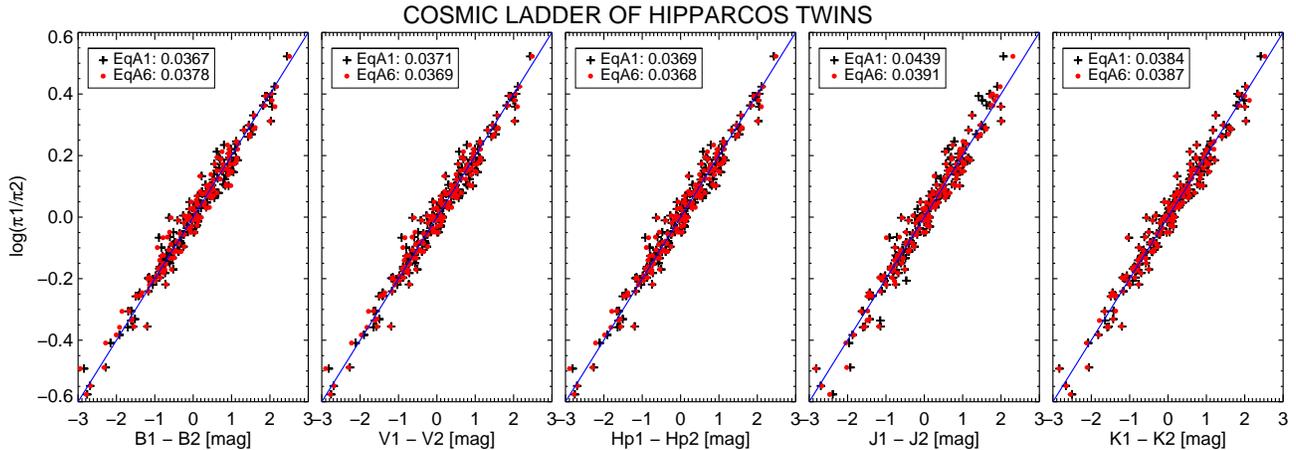}
\caption{Difference of apparent magnitude of twin stars respect to the differences in their Hipparcos parallaxes. Each panel shows the difference for different photometric bands. The red color correspond to the stars when corrected by extinction.  The blue solid line corresponds to the line with slope of 5. The figure shows that Eq.~\ref{plx} is satisfied for twin stars, i.e, the proposition they have the same luminosity is correct.  }
\label{fig:ladder}
\end{center}
\end{figure*}


\section{Results on Distances}\label{results}

\subsection{Twin parallaxes within the  Hipparcos sample}
We considered five different photometric bands to determine the distance from Eq.~(\ref{plx})  for each band.  Only the values having a standard deviation smaller than 20\% of the measured parallax due to different photometric bands were considered to have accurate  photometry, thereby assuring reliable results. These stars are marked with red colour in \fig{fig:hr_twins}. One can see that very few spectroscopic twins yielded inconsistent distances when considering the five different photometric bands.  Uncertainties of the measured parallax of the reference star, as well as the errors in the photometry were considered for calculating the final uncertainty in our parallaxes. This was done using Monte Carlo simulations on the errors in the measured parallax of the reference star and the photometry. We found that the photometric error contributed to negligible uncertainties with respect to the initial parallax error and the standard deviation of the five different photometric bands.  

The mean uncertainty in the Hipparcos parallaxes of our sample is  2.75\%, which propagated to a mean uncertainty in the twin parallaxes  of 3.21\%. The mean error  due the different photometric bands is of 1.72\%, which gives us a mean total uncertainty of 3.46\%  in the twin parallaxes. \\


For the Hipparcos sample, 175  twin pairs with reliable photometry were found. { We show  in \fig{fig:ladder} that our proposition that spectroscopic twin stars have the same luminosity and Eq.~\ref{plx} can be used to determine distances.  {The figure shows the logarithm of the ratio of the twins' parallaxes as measured by Hipparcos,
versus the difference in the apparent magnitudes}. The  {solid line of slope 5 is overplotted with blue}.  Each figure uses the difference of magnitudes in the five different bands considered in this work.  The data has been plotted twice in each panel, the first time we do not consider extinction (essentially Eq.\ref{dist_mod_app}) and the data is plotted in black crosses, while the second one we consider extinction (essentially Eq.\ref{plx}) with red filled circles.  

Our selected spectroscopic twins satisfy our relation very well,  showing like that the difference in the apparent magnitudes is directly related to the difference in their distances, ascending like this the cosmic ladder. 
The figure also tell us that the the corrections due to extinction are normally very small. The corrected magnitudes (red filled circles) have a scatter along the {line of slope 5 that agrees over a range of 3 orders of magnitude} with respect to the uncorrected magnitudes. Also, the scatter is very similar within 3 orders of magnitude for all bands. This is expected as we selected our twins based on stars giving consistent distances between all bands.  We comment here that although we show that extinction is in this case negligible, when applying the method to datasets that expand beyond the local bubble this will not be the case, the general form of Eq.~\ref{plx} should be used.  } 

 \begin{figure*}
\begin{center}
\includegraphics[scale=0.6]{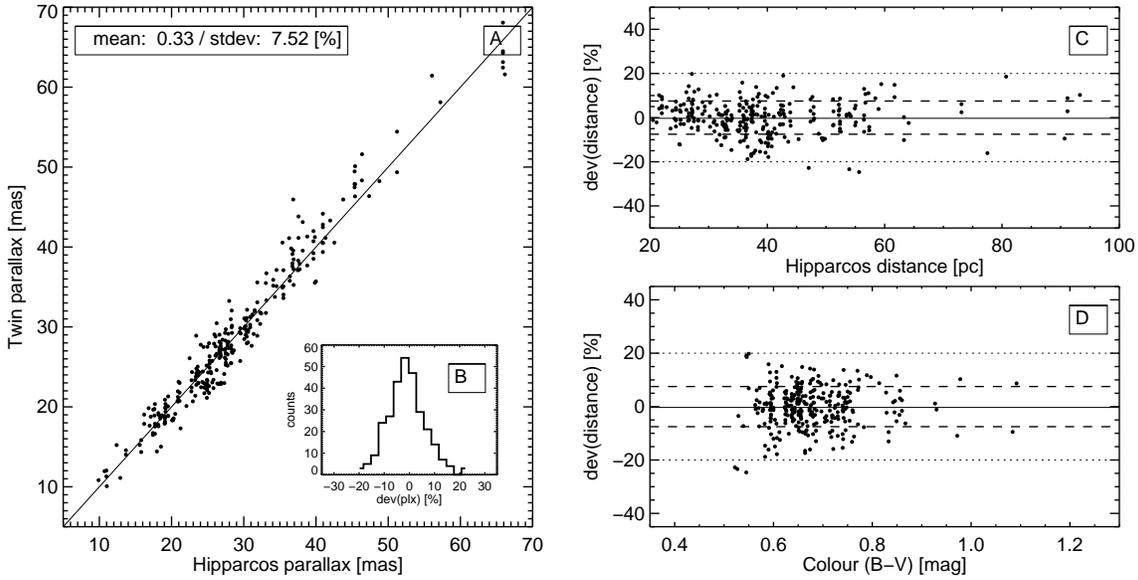}
\caption{Distances obtained with the twin method compared to Hipparcos measurements.  A:  The one-to-one relation is shown with a continuous line and the parallaxes of the twins are shown as dots. B: The distribution of deviations of the parallaxes. C: Deviation of the distance is plotted as a function of distance with dashed line representing  $1 \sigma$ of the deviation distribution of panel B. Dotted line represents the limit of 20\% deviation. D: as panel C  with distances plotted as a function of $B-V$ colour.  }
\label{fig:one-to-one}
\end{center}
\end{figure*}

The comparison of the Hipparcos parallax \citep{2007A&A...474..653V} with our results (twin parallax) are displayed in \fig{fig:one-to-one}.  In Panel A we show the direct comparison of the Hipparcos parallax and the twin parallax for the sample.  The distribution of the deviation\footnote{We define deviation as $\mathrm{dev(plx)} = (1 - \pi_{twin} / \pi_{hip})\times100$ with $\pi_{twin}$ the twin parallax and $\pi_{hip}$ the  Hipparcos measurement.} between our results and Hipparcos is shown in Panel B.  Assuming this distribution can be represented by a Gaussian, its mean and standard deviation are indicated at the top of Panel A.  The agreement  is excellent with a standard deviation of  of 7.5\%.

In Panel C of \fig{fig:one-to-one}  we display the deviation between Hipparcos parallaxes and our results as a function of distance, while in Panel B we display the deviation as a function of colour.  In both panels the dashed lines correspond to the standard deviation of 7.5\%, while the dotted lines represent the deviation of 20\%. We can see that all the determined twins parallaxes in our sample deviate from the Hipparcos ones by less than  25\%.  We can also see that, although our sample is focused on stars that are very close by,  the agreement between our results and those of Hipparcos is independent on the distance.  Similarly, we determine accurate distances for both dwarf and giant stars, i.e. for all colours in our sample. 

Recently, \cite{2015arXiv150105500S} presented a spectroscopic parallax method to determine distance in which they analyse a sample of stars that overlap with our sample. They took approximately 500 main-sequence stars observed with HARPS considering the stellar parameters derived by \cite{2011A&A...533A.141S} and obtained agreements with Hipparcos parallaxes of the order of 20\%. Our results, although providing distances for a subsample, agrees better with 7.5\%.

\subsection{Application to open clusters}\label{clusters}


The stars in open clusters do not have parallax measurements, so our results would mostly compare with the model-dependent method of isochrone fitting to the main-sequence of the open cluster.

To search  for twins in open clusters we had to neglect the second criterion listed in \ref{s:ilis}. The reason is that many of the spectra in the clusters had significantly lower SNR (SNR $\sim$ 15) than the Hipparcos stars, so the measurement of EWs was more uncertain, which produced a larger scatter in fitting the line to the differences in EWs as a function of E.P. We visually inspected the spectra of each of the twin candidates found using the other three criteria listed in \ref{s:ilis}.  The initial sample of twins found in open clusters was reduced with an extra requirement of stars having 2MASS and Johnson-Cousins photometry. We clarify this point here as not all cluster stars initially selected from \cite{2015A&A...577A..47B} had photometry in the four bands required for our present study, in  which we need to compare several photometric bands in the same way as the Hipparcos sample. This requirement is especially important in clusters, where extinction and binary fraction can become significantly larger than in the field.  Only stars for which we obtained a better agreement than 20\% between the distances obtained using the four different photometric bands were selected (cluster stars do not have $H_p$ photometry).  The clusters  M67 and NGC2360 had sufficient member twins of our Hipparcos sample with the required photometric accuracy. Thus, only they were suitable for us to provide a robust value for their distance. For the rest  of the clusters, we either did not find any twin with our library, or we found very few, but most of them lacking  2MASS photometry or not having the radial velocities of the other cluster members. 

It is important to comment  that most of the cluster spectra correspond to giants in the cluster, and our Hipparcos database contains mainly dwarfs. Enlarging the overlap between field and cluster stars can not be easily done since this study was  limited to what is available in the public HARPS dataset. On one hand, most of the field stars are dwarfs because of the intensive search of exoplanets around solar-like stars.  On the other hand, because clusters are more distant, giant stars are bright enough for HARPS to get high SNR. Finding more than 5 twins with good photometry in two clusters using only archive data is indeed very encouraging.  The member stars with twins of Hipparcos are plotted with red colour in the colour-magnitude diagrams of Fig.~\ref{fig:cmd} as taken from WEBDA\footnote{http://www.univie.ac.at/webda/} combined with \cite{2008A&A...489..677P} for the twins found in M67.

 \begin{figure}
\hspace{-1cm}
\includegraphics[scale=0.48]{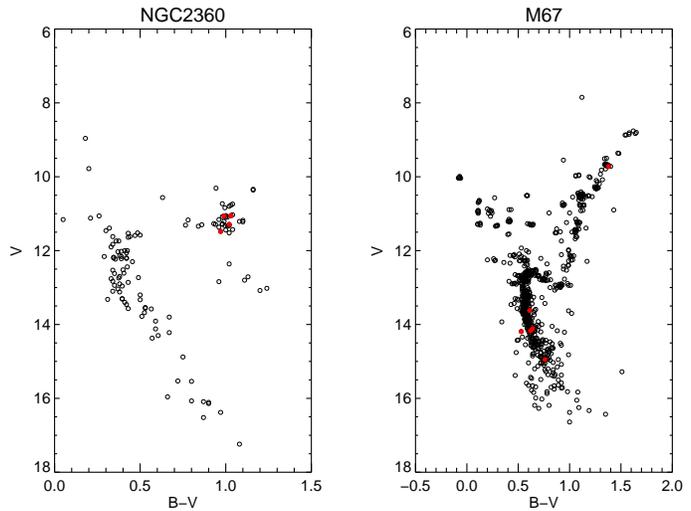}
\caption{Colour-magnitude diagram of the two clusters in this study.  The filled red circles correspond to the star clusters that have a twin in the Hipparcos catalogue. For NGC2360 we only have twins for giant stars, while for M67 we found giants and dwarfs. }
\label{fig:cmd}
\end{figure}

From the diagram we can see that the twins found in NGC2360 are red clump giants while the twins found in M67 are both, main sequence stars and giants. There is in fact only one giant in M67 with an Hipparcos twin, which is not at the red clump. The distances obtained for each of the stars are displayed in \fig{fig:cluster} as a function of colour. The error bars are the square root of the quadratic sum of the error due to the uncertainty of the parallax from the Hipparcos twin and the error due to the distance obtained from the different photometric bands.

 \begin{figure}
\includegraphics[scale=0.45]{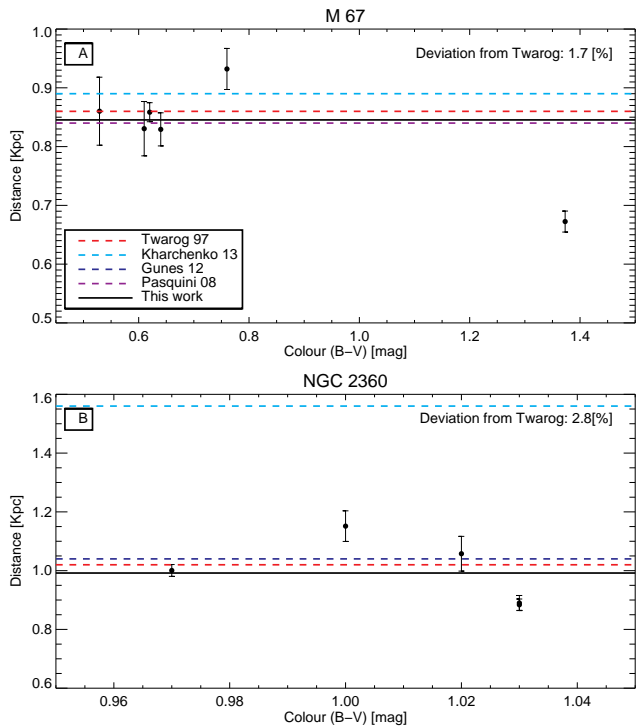}
\caption{Distance of the cluster M67 and NGC2360 in Panel A and B, respectively. The dashed horizontal lines represent values reported in the literature and continuous lines represents our final distance result. Dots correspond to the distances obtained for each twin  found in the cluster, as a function of colour. The error bars represent the standard deviation of the results obtained from different photometric bands. }
\label{fig:cluster}
\end{figure}

The solid line in \fig{fig:cluster} represents the median of the distances obtained for each star individually, the dashed lines represent the values reported in the literature   \citep{1997AJ....114.2556T, 2013A&A...558A..53K, 2012NewA...17..720G, 2008A&A...489..677P}. Not all literature studies listed here report distances for both clusters. \cite{2008A&A...489..677P} determined the distance only for M67 while \cite{2012NewA...17..720G} only for NGC2360.  In \fig{fig:cluster} we show in the legend the agreement with \cite{1997AJ....114.2556T}, who analysed both clusters and agrees with us better than 3\%. Note that all literature values except that one of  \cite{2008A&A...489..677P} determined the distance using evolutionary models by fitting the main-sequence to the colour-magnitude diagram of the cluster. Our method is more similar to \cite{2008A&A...489..677P}, which considered the averaged magnitude of ten solar analogs in M67 and the absolute magnitude of the Sun to compute the distance modulus. The agreement with \cite{2008A&A...489..677P}   is remarkable. Also note that the discrepancy in distances for NGC2360 using main-sequence fitting exceeds 35\% for the case of  \cite{2013A&A...558A..53K} respect to the other works. 

It is  curious that our result obtained from the reddest (giant) star in M67 is different than those obtained from the bluer (main-sequence) stars. Although we did not find an offset for giant stars in our Hipparcos sample, measuring distances to red giants can be very challenging \citep[e.g.][]{2013MNRAS.429.3645S, 2014MNRAS.437..351B}.  The shape of the red giant branch (\fig{fig:cmd}), demonstrates how a slight change in  colour (temperature) implies a large change in magnitude (luminosity).  Having a larger sample of giants is needed to understand why this star gave a lower distance than its dwarfs siblings in M67. Such insights could be revealed from a stellar survey like APOGEE or Gaia-ESO, where large samples of giants  are observed in clusters and in the field.   \\

\section{Discussion and conclusions} \label{fin}
The {twin method} is simple, and is based on two fundamental astrophysical principles: the apparent magnitude of a star decreases with distance; and if two stars are equivalent then their spectra are identical. When several photometric bands give consistent magnitudes and the parallax of the reference star is well measured, the { twin method} has  important applications in stellar and Galactic astrophysics beyond the direct determination of distances. 

It becomes a powerful method to  complement future Gaia parallaxes. It also allows in-depth studies into astrophysical mechanisms affecting stellar luminosities.   
For example, Fig.~\ref{fig:one-to-one} shows that there are cases with distances that deviate more than 15\% from Hipparcos. Our twins have been found under the assumption that the flux around the spectral lines of 11 chemical elements agree. It is possible that the absorption produced by other elements, such as carbon or neutron-capture elements, might change opacities such  that the evolution, and thus the luminosity,  is slightly affected.  In our procedure to find twins, every EWs has the same weight when we perform the fitting of the trends. Other elements that are not included in the selection of lines of the Gaia FGK benchmark stars in \cite[][2015]{2014A&A...564A.133J} are less numerous in the optical spectra.  Including the EWs of some lines from these elements would produce an effectively unaffected fit of the trends.  The study of the effect of variation of other elements not listed in \cite[][2015]{2014A&A...564A.133J} in the deviation of the distance goes beyond the purpose of the current paper, which consists in introducing the concept of using twin stars for the determination of distances. We have shown here that by comparing equivalent widths of the elements Mg, Si, Ca, Ti, Sc, V, Cr, Mn, Fe, Co and Ni we are able to find stars that are equal enough to determine distances within 7.5\% agreement with Hipparcos parallaxes.

The {twin method} also provides an attractive approach to quantify the effects of dust extinction in the Milky Way. The factor $\mathrm{R}(X)$ used in Eq.~(\ref{plx}) is assumed to be constant in the disk, but this value originates from calibrations \citep{1999PASP..111...63F, 2013MNRAS.430.2188Y}. One can refine  these calibrations  by identifying twins with good photometry and parallaxes that are located in different parts of the Galaxy. Furthermore, one can test if $\mathrm{R}(X)$ is still constant outside the disk.   Such analysis could be done when Gaia releases its parallaxes, giving us accurate distances for stars beyond the solar vicinity and the disk. 


It is worth to comment that our method, by being restricted to twin stars, provides distances to a much smaller number of stars. This happens however to several other methods that are also restricted to particular kind of stars, such as distances to variable stars \citep[e.g.][]{2013A&A...554A.132G} or distances using asteroseismic data \citep{2013AN....334...22S}. In this era of large datasets of stellar spectra and with Gaia parallaxes coming soon, samples of twin pairs with one star having a parallax will become numerous in the near future.  

With the relatively small sample of 536 high resolution spectra of FGK stars we found 175 twin pairs.  We showed that with accurate photometry and parallaxes we can retrieve twin parallaxes with typical errors of 3.25\%, which is of the order of the Hipparcos uncertainties for our sample. Our results agree within 7.5\% for most of the cases with the measurements of Hipparcos.  From this sample we also determined the distances for the clusters M67 and NGC2360, finding agreement of  $\sim 3$\%  with standard techniques such as that one of \cite{1997AJ....114.2556T}.  { Our main result  can be seen in \fig{fig:ladder}, which demonstrates that our basic proposition, that twin stars have the same luminosity
(i.e. absolute magnitude) {is verified}. }

Currently, high-resolution spectra of Hipparcos stars can be found in several public archives. Extending our sample of reference stars will allow us to find more twins of Hipparcos in on-going high-resolution spectroscopic surveys such as Gaia-ESO, APOGEE or GALAH, and determine distances of  fainter objects that do not benefit from direct parallaxes. With Gaia measuring the most accurate parallaxes in history, millions of stars already observed with high-resolution spectra will soon have accurate distances, allowing us to climb the cosmic distance ladder.  Indeed, with the next generation of large telescopes, such as E-ELT, high resolution spectra of objects beyond the reach of Gaia by several magnitudes, will become available.  We look forward to applying the {twin method} to stars too distant for accurate parallaxes from Gaia, but for which good spectra can be obtained with large telescopes, and helping to understand fundamental astrophysical processes from the deepest parts of our Galaxy.

\subsubsection*{Acknowledgments}
It is our pleasure to thank T. Masseron, K. Hawkins and J. Penner for fruitful discussions on the subject.  { The authors thank Chris Flynn for the good suggestions made in the referee report, which helped to improve the paper significantly. }
This work was partly supported by the European Union FP7 programme through ERC grant number 320360. C.~W. acknowledges Leverhulme Trust through grant RPG-2012-541.  
Based on data obtained from the ESO Science Archive Facility.
This research has made use of the SIMBAD and WEBDA database. 
 P.~J. and T.~M. thank Damian and Kian for motivating this work.

\bibliography{refs_twins}

\begin{thebibliography}{56}
\expandafter\ifx\csname natexlab\endcsname\relax\def\natexlab#1{#1}\fi

\bibitem[{{Allende Prieto} {et~al}\mbox{.}(2008){Allende Prieto}, {Majewski},
  {Schiavon}, {Cunha}, {Frinchaboy}, {Holtzman}, {Johnston}, {Shetrone},
  {Skrutskie}, {Smith}, \& {Wilson}}]{2008AN....329.1018A}
{Allende Prieto} C. {et~al.}, 2008, Astronomische Nachrichten, 329, 1018

\bibitem[{{Alonso}, {Arribas} \& {Mart{\'{\i}}nez-Roger}(1999){Alonso},
  {Arribas}, \& {Mart{\'{\i}}nez-Roger}}]{1999A&AS..140..261A}
{Alonso} A., {Arribas} S., {Mart{\'{\i}}nez-Roger} C., 1999, \aaps, 140, 261

\bibitem[{{Binney} {et~al}\mbox{.}(2014){Binney}, {Burnett}, {Kordopatis},
  {McMillan}, {Sharma}, {Zwitter}, {Bienaym{\'e}}, {Bland-Hawthorn},
  {Steinmetz}, {Gilmore}, {Williams}, {Navarro}, \&
  {Grebel}}]{2014MNRAS.437..351B}
{Binney} J. {et~al.}, 2014, MNRAS, 437, 351

\bibitem[{{Blanco-Cuaresma} {et~al}\mbox{.}(2015){Blanco-Cuaresma}, {Soubiran},
  {Heiter}, {Asplund}, {Carraro}, {Costado}, {Feltzing},
  {Gonz{\'a}lez-Hern{\'a}ndez}, {Jim{\'e}nez-Esteban}, {Korn}, {Marino},
  {Montes}, {San Roman}, {Tabernero}, \& {Tautvai{\v
  s}ien{\.e}}}]{2015A&A...577A..47B}
{Blanco-Cuaresma} S. {et~al.}, 2015, \aap, 577, A47

\bibitem[{{Blanco-Cuaresma}
  {et~al}\mbox{.}(2014{\natexlab{a}}){Blanco-Cuaresma}, {Soubiran}, {Heiter},
  \& {Jofr{\'e}}}]{2014A&A...569A.111B}
{Blanco-Cuaresma} S., {Soubiran} C., {Heiter} U., {Jofr{\'e}} P.,
  2014{\natexlab{a}}, A\&A, 569, A111

\bibitem[{{Blanco-Cuaresma}
  {et~al}\mbox{.}(2014{\natexlab{b}}){Blanco-Cuaresma}, {Soubiran},
  {Jofr{\'e}}, \& {Heiter}}]{2014A&A...566A..98B}
{Blanco-Cuaresma} S., {Soubiran} C., {Jofr{\'e}} P., {Heiter} U.,
  2014{\natexlab{b}}, A\&A, 566, A98

\bibitem[{{Burnett} \& {Binney}(2010)}]{2010MNRAS.407..339B}
{Burnett} B., {Binney} J., 2010, \mnras, 407, 339

\bibitem[{{Casey}(2014)}]{2014arXiv1405.5968C}
{Casey} A.~R., 2014, ArXiv e-prints

\bibitem[{{Cassisi}(2014)}]{2014EAS....65...17C}
{Cassisi} S., 2014, in EAS Publications Series, Vol.~65, , pp. 17--74

\bibitem[{{Datson}, {Flynn} \& {Portinari}(2014){Datson}, {Flynn}, \&
  {Portinari}}]{2014MNRAS.439.1028D}
{Datson} J., {Flynn} C., {Portinari} L., 2014, \mnras, 439, 1028

\bibitem[{{de Jong}(2011)}]{2011Msngr.145...14J}
{de Jong} R., 2011, The Messenger, 145, 14

\bibitem[{{De Pascale} {et~al}\mbox{.}(2014){De Pascale}, {Worley}, {de
  Laverny}, {Recio-Blanco}, {Hill}, \& {Bijaoui}}]{2014A&A...570A..68D}
{De Pascale} M., {Worley} C.~C., {de Laverny} P., {Recio-Blanco} A., {Hill} V.,
  {Bijaoui} A., 2014, A\&A, 570, A68

\bibitem[{{De Silva} {et~al}\mbox{.}(2015){De Silva}, {Freeman},
  {Bland-Hawthorn}, {Martell}, {de Boer}, {Asplund}, \&
  {Keller}}]{2015arXiv150204767D}
{De Silva} G.~M., {Freeman} K.~C., {Bland-Hawthorn} J., {Martell} S., {de Boer}
  E.~W., {Asplund} M., {Keller} S., 2015, ArXiv e-prints

\bibitem[{{ESA}(1997)}]{1997ESASP1200.....E}
{ESA}, ed., 1997, ESA Special Publication, Vol. 1200, {The HIPPARCOS and TYCHO
  catalogues. Astrometric and photometric star catalogues derived from the ESA
  HIPPARCOS Space Astrometry Mission}

\bibitem[{{Fitzpatrick}(1999)}]{1999PASP..111...63F}
{Fitzpatrick} E.~L., 1999, PASP, 111, 63

\bibitem[{{Genovali} {et~al}\mbox{.}(2013){Genovali}, {Lemasle}, {Bono},
  {Romaniello}, {Primas}, {Fabrizio}, {Buonanno}, {Fran{\c c}ois}, {Inno},
  {Laney}, {Matsunaga}, {Pedicelli}, \& {Th{\'e}venin}}]{2013A&A...554A.132G}
{Genovali} K. {et~al.}, 2013, \aap, 554, A132

\bibitem[{{Gilmore} {et~al}\mbox{.}(2012){Gilmore}, {Randich}, {Asplund},
  {Binney}, {Bonifacio}, {Drew}, {Feltzing}, \&
  {Ferguson}}]{2012Msngr.147...25G}
{Gilmore} G., {Randich} S., {Asplund} M., {Binney} J., {Bonifacio} P., {Drew}
  J., {Feltzing} S., {Ferguson} A., 2012, The Messenger, 147, 25

\bibitem[{{Gratton} {et~al}\mbox{.}(2003){Gratton}, {Bragaglia}, {Carretta},
  {Clementini}, {Desidera}, {Grundahl}, \& {Lucatello}}]{2003A&A...408..529G}
{Gratton} R.~G., {Bragaglia} A., {Carretta} E., {Clementini} G., {Desidera} S.,
  {Grundahl} F., {Lucatello} S., 2003, A\&A, 408, 529

\bibitem[{{Gray}(1992)}]{1992oasp.book.....G}
{Gray} D.~F., 1992, {The observation and analysis of stellar photospheres.}

\bibitem[{{Groenewegen}(2008)}]{2008A&A...488..935G}
{Groenewegen} M.~A.~T., 2008, A\&A, 488, 935

\bibitem[{{Groenewegen}, {Udalski} \& {Bono}(2008){Groenewegen}, {Udalski}, \&
  {Bono}}]{2008A&A...481..441G}
{Groenewegen} M.~A.~T., {Udalski} A., {Bono} G., 2008, A\&A, 481, 441

\bibitem[{{G{\"u}ne{\c s}}, {Karata{\c s}} \& {Bonatto}(2012){G{\"u}ne{\c s}},
  {Karata{\c s}}, \& {Bonatto}}]{2012NewA...17..720G}
{G{\"u}ne{\c s}} O., {Karata{\c s}} Y., {Bonatto} C., 2012, New, 17, 720

\bibitem[{{Iben}(1967{\natexlab{a}})}]{1967Sci...155..785I}
{Iben}, Jr. I., 1967{\natexlab{a}}, Science, 155, 785

\bibitem[{{Iben}(1967{\natexlab{b}})}]{1967ApJ...147..624I}
{Iben}, Jr. I., 1967{\natexlab{b}}, \apj, 147, 624

\bibitem[{{Ivezi{\'c}} {et~al}\mbox{.}(2008){Ivezi{\'c}}, {Sesar}, {Juri{\'c}},
  {Bond}, {Dalcanton}, {Rockosi}, {Yanny}, {Newberg}, {Beers}, {Allende
  Prieto}, {Wilhelm}, {Lee}, {Sivarani}, {Norris}, {Bailer-Jones}, {Re
  Fiorentin}, {Schlegel}, {Uomoto}, {Lupton}, {Knapp}, {Gunn}, {Covey},
  {Smith}, {Miknaitis}, {Doi}, {Tanaka}, {Fukugita}, {Kent}, {Finkbeiner},
  {Munn}, {Pier}, {Quinn}, {Hawley}, {Anderson}, {Kiuchi}, {Chen}, {Bushong},
  {Sohi}, {Haggard}, {Kimball}, {Barentine}, {Brewington}, {Harvanek},
  {Kleinman}, {Krzesinski}, {Long}, {Nitta}, {Snedden}, {Lee}, {Harris},
  {Brinkmann}, {Schneider}, \& {York}}]{2008ApJ...684..287I}
{Ivezi{\'c}} {\v Z}. {et~al.}, 2008, \apj, 684, 287

\bibitem[{{Jofr{\'e}} {et~al}\mbox{.}(2014{\natexlab{a}}){Jofr{\'e}}, {Heiter},
  {Blanco-Cuaresma}, \& {Soubiran}}]{2014ASInC..11..159J}
{Jofr{\'e}} P., {Heiter} U., {Blanco-Cuaresma} S., {Soubiran} C.,
  2014{\natexlab{a}}, in Astronomical Society of India Conference Series,
  Vol.~11, Astronomical Society of India Conference Series, pp. 159--166

\bibitem[{{Jofr{\'e}} {et~al}\mbox{.}(2015){Jofr{\'e}}, {Heiter}, {Soubiran},
  {Blanco-Cuaresma}, {Masseron}, {Nordlander}, {Chemin}, {Worley}, {Van Eck},
  {Hourihane}, {Gilmore}, {Adibekyan}, {Bergemann}, {Cantat-Gaudin},
  {Delgado-Mena}, {Gonz{\'a}lez Hern{\'a}ndez}, {Guiglion}, {Lardo}, {de
  Laverny}, {Lind}, {Magrini}, {Mikolaitis}, {Montes}, {Pancino},
  {Recio-Blanco}, {Sordo}, {Sousa}, {Tabernero}, \&
  {Vallenari}}]{2015arXiv150700027J}
{Jofr{\'e}} P. {et~al.}, 2015, ArXiv e-prints

\bibitem[{{Jofr{\'e}} {et~al}\mbox{.}(2014{\natexlab{b}}){Jofr{\'e}}, {Heiter},
  {Soubiran}, {Blanco-Cuaresma}, {Worley}, {Pancino}, {Cantat-Gaudin},
  {Magrini}, {Bergemann}, {Gonz{\'a}lez Hern{\'a}ndez}, {Hill}, {Lardo}, {de
  Laverny}, {Lind}, {Masseron}, {Montes}, {Mucciarelli}, {Nordlander}, {Recio
  Blanco}, {Sobeck}, {Sordo}, {Sousa}, {Tabernero}, {Vallenari}, \& {Van
  Eck}}]{2014A&A...564A.133J}
{Jofr{\'e}} P. {et~al.}, 2014{\natexlab{b}}, A\&A, 564, A133

\bibitem[{{Katz} {et~al}\mbox{.}(2011){Katz}, {Soubiran}, {Cayrel}, {Barbuy},
  {Friel}, {Bienaym{\'e}}, \& {Perrin}}]{2011A&A...525A..90K}
{Katz} D., {Soubiran} C., {Cayrel} R., {Barbuy} B., {Friel} E., {Bienaym{\'e}}
  O., {Perrin} M.-N., 2011, A\&A, 525, A90

\bibitem[{{Kharchenko} \& {\it et al.}(2013)}]{2013A&A...558A..53K}
{Kharchenko} N.~V., {\it et al.}, 2013, A\&A, 558, A53

\bibitem[{{Lallement} {et~al}\mbox{.}(2003){Lallement}, {Welsh}, {Vergely},
  {Crifo}, \& {Sfeir}}]{2003A&A...411..447L}
{Lallement} R., {Welsh} B.~Y., {Vergely} J.~L., {Crifo} F., {Sfeir} D., 2003,
  \aap, 411, 447

\bibitem[{{Mayor} {et~al}\mbox{.}(2003){Mayor}, {Pepe}, {Queloz}, {Bouchy},
  {Rupprecht}, {Lo Curto}, {Avila}, \& {Benz}}]{2003Msngr.114...20M}
{Mayor} M., {Pepe} F., {Queloz} D., {Bouchy} F., {Rupprecht} G., {Lo Curto} G.,
  {Avila} G., {Benz} W., 2003, The Messenger, 114, 20

\bibitem[{{Mel{\'e}ndez}, {Dodds-Eden} \& {Robles}(2006){Mel{\'e}ndez},
  {Dodds-Eden}, \& {Robles}}]{2006ApJ...641L.133M}
{Mel{\'e}ndez} J., {Dodds-Eden} K., {Robles} J.~A., 2006, ApJL, 641, L133

\bibitem[{{Nidever} {et~al}\mbox{.}(2014){Nidever}, {Bovy}, {Bird}, {Andrews},
  {Hayden}, {Holtzman}, {Majewski}, \& {Smith}}]{2014ApJ...796...38N}
{Nidever} D.~L., {Bovy} J., {Bird} J.~C., {Andrews} B.~H., {Hayden} M.,
  {Holtzman} J., {Majewski} S.~R., {Smith} V., 2014, ApJ, 796, 38

\bibitem[{{Pasquini} {et~al}\mbox{.}(2008){Pasquini}, {Biazzo}, {Bonifacio},
  {Randich}, \& {Bedin}}]{2008A&A...489..677P}
{Pasquini} L., {Biazzo} K., {Bonifacio} P., {Randich} S., {Bedin} L.~R., 2008,
  A\&A, 489, 677

\bibitem[{{Perryman} {et~al}\mbox{.}(2001){Perryman}, {de Boer}, {Gilmore},
  {H{\o}g}, {Lattanzi}, {Lindegren}, {Luri}, {Mignard}, {Pace}, \& {de
  Zeeuw}}]{2001A&A...369..339P}
{Perryman} M.~A.~C. {et~al.}, 2001, A\&A, 369, 339

\bibitem[{{Perryman} {et~al}\mbox{.}(1997){Perryman}, {Lindegren},
  {Kovalevsky}, {Hoeg}, {Bastian}, {Bernacca}, {Cr{\'e}z{\'e}}, {Donati},
  {Grenon}, {Grewing}, {van Leeuwen}, {van der Marel}, {Mignard}, {Murray}, {Le
  Poole}, {Schrijver}, {Turon}, {Arenou}, {Froeschl{\'e}}, \&
  {Petersen}}]{1997A&A...323L..49P}
{Perryman} M.~A.~C. {et~al.}, 1997, A\&A, 323, L49

\bibitem[{{Ram{\'{\i}}rez} {et~al}\mbox{.}(2014){Ram{\'{\i}}rez},
  {Mel{\'e}ndez}, {Bean}, {Asplund}, {Bedell}, {Monroe}, {Casagrande},
  {Schirbel}, {Dreizler}, {Teske}, {Tucci Maia}, {Alves-Brito}, \&
  {Baumann}}]{2014A&A...572A..48R}
{Ram{\'{\i}}rez} I. {et~al.}, 2014, A\&A, 572, A48

\bibitem[{{Rodrigues} {et~al}\mbox{.}(2014){Rodrigues}, {Girardi}, {Miglio},
  {Bossini}, {Bovy}, {Epstein}, {Pinsonneault}, {Stello}, {Zasowski}, {Prieto},
  \& {Chaplin}}]{2014MNRAS.445.2758R}
{Rodrigues} T.~S. {et~al.}, 2014, MNRAS, 445, 2758

\bibitem[{{Santiago} {et~al}\mbox{.}(2015){Santiago}, {Brauer}, {Anders},
  {Chiappini}, {Girardi}, {Rocha-Pinto}, {Balbinot}, {da Costa}, {Maia},
  {Schultheis}, {Steinmetz}, {Miglio}, {Montalb{\'a}n}, {Schneider}, {Beers},
  {Frinchaboy}, {Lee}, \& {Zasowski}}]{2015arXiv150105500S}
{Santiago} B.~X. {et~al.}, 2015, ArXiv e-prints

\bibitem[{{Schlafly} \& {Finkbeiner}(2011)}]{2011ApJ...737..103S}
{Schlafly} E.~F., {Finkbeiner} D.~P., 2011, \apj, 737, 103

\bibitem[{{Schlegel}, {Finkbeiner} \& {Davis}(1998){Schlegel}, {Finkbeiner}, \&
  {Davis}}]{1998ApJ...500..525S}
{Schlegel} D.~J., {Finkbeiner} D.~P., {Davis} M., 1998, \apj, 500, 525

\bibitem[{{Schultheis} {et~al}\mbox{.}(2014){Schultheis}, {Zasowski}, {Allende
  Prieto}, {Anders}, {Beaton}, {Beers}, {Bizyaev}, {Chiappini}, {Frinchaboy},
  {Garc{\'{\i}}a P{\'e}rez}, {Ge}, {Hearty}, {Holtzman}, {Majewski}, {Muna},
  {Nidever}, {Shetrone}, \& {Schneider}}]{2014AJ....148...24S}
{Schultheis} M. {et~al.}, 2014, \aj, 148, 24

\bibitem[{{Serenelli} {et~al}\mbox{.}(2013){Serenelli}, {Bergemann}, {Ruchti},
  \& {Casagrande}}]{2013MNRAS.429.3645S}
{Serenelli} A.~M., {Bergemann} M., {Ruchti} G., {Casagrande} L., 2013, \mnras,
  429, 3645

\bibitem[{{Silva Aguirre} {et~al}\mbox{.}(2013){Silva Aguirre}, {Casagrande},
  {Basu}, {Campante}, {Chaplin}, {Huber}, {Miglio}, \&
  {Serenelli}}]{2013AN....334...22S}
{Silva Aguirre} V., {Casagrande} L., {Basu} S., {Campante} T.~L., {Chaplin}
  W.~J., {Huber} D., {Miglio} A., {Serenelli} A.~M., 2013, Astronomische
  Nachrichten, 334, 22

\bibitem[{{Skrutskie} {et~al}\mbox{.}(2006){Skrutskie}, {Cutri}, {Stiening},
  {Weinberg}, {Schneider}, \& {Carpenter}}]{2006AJ....131.1163S}
{Skrutskie} M.~F., {Cutri} R.~M., {Stiening} R., {Weinberg} M.~D., {Schneider}
  S., {Carpenter} J.~M., 2006, AJ, 131, 1163

\bibitem[{{Soubiran} {et~al}\mbox{.}(2008){Soubiran}, {Bienaym{\'e}},
  {Mishenina}, \& {Kovtyukh}}]{2008A&A...480...91S}
{Soubiran} C., {Bienaym{\'e}} O., {Mishenina} T.~V., {Kovtyukh} V.~V., 2008,
  \aap, 480, 91

\bibitem[{{Soubiran}, {Bienaym{\'e}} \& {Siebert}(2003){Soubiran},
  {Bienaym{\'e}}, \& {Siebert}}]{2003A&A...398..141S}
{Soubiran} C., {Bienaym{\'e}} O., {Siebert} A., 2003, A\&A, 398, 141

\bibitem[{{Soubiran} {et~al}\mbox{.}(2010){Soubiran}, {Le Campion}, {Cayrel de
  Strobel}, \& {Caillo}}]{2010A&A...515A.111S}
{Soubiran} C., {Le Campion} J.-F., {Cayrel de Strobel} G., {Caillo} A., 2010,
  A\&A, 515, A111

\bibitem[{{Sousa} {et~al}\mbox{.}(2011){Sousa}, {Santos}, {Israelian}, {Mayor},
  \& {Udry}}]{2011A&A...533A.141S}
{Sousa} S.~G., {Santos} N.~C., {Israelian} G., {Mayor} M., {Udry} S., 2011,
  \aap, 533, A141

\bibitem[{{Steinmetz} {et~al}\mbox{.}(2006){Steinmetz}, {Zwitter}, {Siebert},
  {Watson}, {Freeman}, {Munari}, {Campbell}, \&
  {Williams}}]{2006AJ....132.1645S}
{Steinmetz} M., {Zwitter} T., {Siebert} A., {Watson} F.~G., {Freeman} K.~C.,
  {Munari} U., {Campbell} R., {Williams} M., 2006, AJ, 132, 1645

\bibitem[{{Twarog}, {Ashman} \& {Anthony-Twarog}(1997){Twarog}, {Ashman}, \&
  {Anthony-Twarog}}]{1997AJ....114.2556T}
{Twarog} B.~A., {Ashman} K.~M., {Anthony-Twarog} B.~J., 1997, AJ, 114, 2556

\bibitem[{{van Leeuwen}(2007)}]{2007A&A...474..653V}
{van Leeuwen} F., 2007, A\&A, 474, 653

\bibitem[{{Wenger} {et~al}\mbox{.}(2000){Wenger}, {Ochsenbein}, {Egret},
  {Dubois}, {Bonnarel}, {Borde}, {Genova}, {Jasniewicz}, {Lalo{\"e}},
  {Lesteven}, \& {Monier}}]{2000A&AS..143....9W}
{Wenger} M. {et~al.}, 2000, \aaps, 143, 9

\bibitem[{{Xue} {et~al}\mbox{.}(2014){Xue}, {Ma}, {Rix}, {Morrison}, {Harding},
  {Beers}, \& {Ivans}}]{2014ApJ...784..170X}
{Xue} X.-X., {Ma} Z., {Rix} H.-W., {Morrison} H.~L., {Harding} P., {Beers}
  T.~C., {Ivans} I.~I., 2014, ApJ, 784, 170

\bibitem[{{Yuan}, {Liu} \& {Xiang}(2013){Yuan}, {Liu}, \&
  {Xiang}}]{2013MNRAS.430.2188Y}
{Yuan} H.~B., {Liu} X.~W., {Xiang} M.~S., 2013, MNRAS, 430, 2188

\end{thebibliography}

\begin{appendix}
\section{Formalism of distance determination  of twin stars}\label{formalism}

By restricting ourself to use twin stars, we can apply the classical expression described in astronomy textbooks to measure stellar distances  {\it directly}. 
Suppose there are two  stars in the Galaxy  at different locations. 
Assuming star $1$ and star $2$  are  twins, they have the same luminosity. Thus their respective apparent magnitudes $m_1$ and $m_2$ can be related to their distances $d_1$ and $d_2$ via the relation
\begin{equation}\label{dist_mod_app}
m_{1} - m_{2} = -5 \log (d_1/d_2)\;.
\end{equation}
Depending on the location and distance of each star in the Galaxy, their apparent brightnesses will suffer from interstellar extinction. Suppose both stars  are observed with an arbitrary  filter $X$,  their apparent magnitudes $m_i\;,(i=1,2),$ may assumed to be
\begin{equation}
m_i = X_i - \mathrm{A}(X)_i\;,
\end{equation}
where $\mathrm{A}(X)_i$ represents the correction due to extinction in the interstellar medium, may be expressed as function of the colour $X-Y$. 
\begin{equation}\label{ext_app}
\mathrm{A}(X) = \mathrm{R}(X)\mathrm{E}(X-Y)\;,
\end{equation}
where $\mathrm{R}(X)$ is the ratio of the of total-to-selective extinction in filter $X$,  and $\mathrm{E}(X-Y)$ is the de-reddening of the stars' colour \citep[e.g.][]{1999PASP..111...63F}.  
The value  of the ratio of the of total-to-selective extinction $\mathrm{R}(X)$ may be assumed constant in the Galactic disk \citep[e.g.][]{1999PASP..111...63F, 2013MNRAS.430.2188Y,  2011ApJ...737..103S}, 
therefore Eq.~(\ref{dist_mod_app}) becomes
\begin{equation}\label{dist_ext_appE}
X_1- X_2-\mathrm{R}(X) \Big[\mathrm{E}(X-Y)_1 -  \mathrm{E}(X-Y)_2\Big] = -5 \log (d_1/d_2)\;.
\end{equation}
It is seen that the difference of the logarithm of distances becomes a function of the difference of the apparent brightnesses and the difference of the de-reddening of both stars.   
Since the de-reddening is a correction to the intrinsic colour $(X-Y)_0$ of the stars,
and since the stars are twins, their intrinsic colour is the same,  which implies 
\begin{equation}\label{twin_color}
 \mathrm{E}(X-Y)_1 - \mathrm{E}(X-Y)_2 =(X-Y)_1 -  (X-Y)_2 \;.
\end{equation}
Hence,  for twins stars the difference in their observed colours is the same as the difference in their de-reddening.  This is important because by considering the colours we do not need to consider de-reddening factors from extinction maps or calibrations of extinction in different photometric bands, which would bring external uncertainties to our analysis.  Using Eq.~(\ref{twin_color}),  Eq.~(\ref{dist_ext_appE})  becomes our final expression 
\begin{equation}\label{plx}
X_1 - {X_2} -\mathrm{R}(X) \Big [(Y-X)_1 - (Y-X)_2\Big]  =  -5 \log (d_1/d_2)\;. 
\end{equation}

 This formula can be used to determine the distance of one star of the twin pair  knowing the photometry of both of them, the value of $\mathrm{R}(X)$, and the distance of its twin counterpart. Note that the only dependency on extinction in this formula is the factor $\mathrm{R}(X)$. 
 { The demonstration of this formula can be seen in \fig{fig:ladder} with a sample of  Hipparcos twin stars, {in which we verify that in
a log-log plot, the data lie on a line with slope 5}.  In the figure we also plot the relation for Eq.~\ref{dist_mod_app} and we see that the extinction correction is negligible. This is expected since the Hipparcos twin stars lie in the local bubble.  This is not the case for stars beyond the local bubble, for which the correction term of Eq.~\ref{plx} should be used.}

 It is worth commenting here that the factor $\mathrm{R}(X)$ is indirectly used by any other method determining distances based on apparent magnitudes of stars. Usually, to know the absolute magnitude of an individual star (or its distance modulus) would require the correction of the observed magnitude from extinction, which is done 
 applying Eq.~(\ref{ext_app}) to the magnitudes. The factor $\mathrm{E}(X-Y)$ is commonly taken from extinction maps, such as those from \cite{1998ApJ...500..525S} and \cite{2011ApJ...737..103S} for the $(B-V)$ colour. Extinction maps are different from each other \citep[see][for such analysis]{2014AJ....148...24S}, as well as the empirical corrections applied to $\mathrm{E}(B-V)$ for correcting extinction of magnitudes taken with other photometric filters \citep[e.g.][]{2013MNRAS.430.2188Y}.   Extinction coefficients can also be determined together with the distance indirectly \citep[e.g.][]{2015arXiv150105500S}. By using the differential magnitudes in different photometric bands as well as the fact that twins have intrinsically the same absolute magnitudes and colours, our expression to determine distances (Eq.~\ref{plx}) is independent on extinction maps or coefficients.

\section{Stellar atmospheric parameters}
 As a consistency check, we compared the stellar parameters (effective temperature, surface gravity, metallicity and abundance of $\alpha$ elements with respect to iron) of our twins as obtained within the AMBRE project for the HARPS dataset \citep{2014A&A...570A..68D}, for which 135 pairs had both stars with well determined parameters. The remaining stars either the AMBRE method to determine the parameters did not converge to a physical solution, or our stars were observed after the AMBRE project collected the data. The atmospheric  parameters of the  twin sample are shown in Fig.~\ref{fig:ambre}, where in the x-axis we plot the parameters of the first star and in the y-axis the parameters of the twin.   The agreement is excellent, with negligible offsets and standard deviation that are smaller than typical errors on the stellar parameters.  From this test we note the following:

 \begin{figure*}
\begin{center}
\includegraphics[scale=0.65]{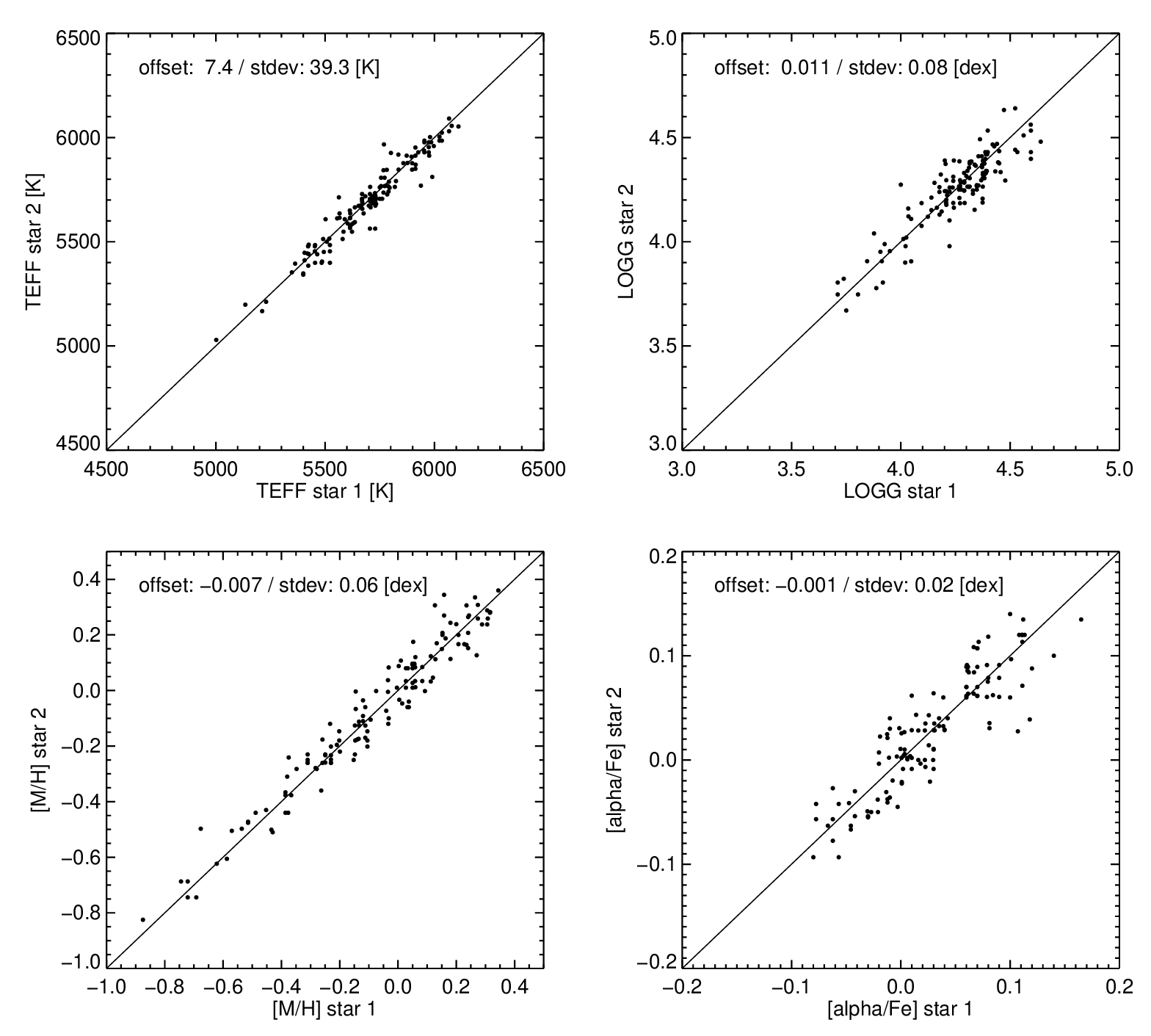}
\vspace{0.5cm}
\caption{Comparison of effective temperatures, surface gravities, metallicities and $\alpha$ abundances for the twin pairs found in the HARPS sample. The parameters were determined homogeneously and automatically as part of the AMBRE project.}
\label{fig:ambre}
\end{center}
\end{figure*}

\begin{itemize}
\item For twin stars the  AMBRE parameters agree remarkably well, confirming the good performance of this method to determine parameters. 
\item Our method yields consistent results with standard methods to determine stellar parameters. This validates the concept that a pure comparison of observed spectra serves to determine parameters, which has been already shown by \cite{2003A&A...398..141S}. 
This also validates that the 11 elements and 423 atomic lines chosen for the comparison are a good proxy for the determination of stellar parameters. 
\item Retrieving surface gravities more accurately than 0.1~dex from pure spectroscopy is extremely challenging. Even in this case, where we have high signal-to-noise and high-resolution spectra that are practically identical, we find differences in the AMBRE gravities of $\sim0.3$~dex or more. 
\item According to AMBRE parameters, our twin stars span a wide range of temperatures and metallicities, although they are all concentrated towards high gravities.  The reason is probably that our initial sample contains mostly main-sequence stars (see Fig.~\ref{fig:hr}). 
 Nonetheless,  from the HR diagram in Fig.~\ref{fig:hr_twins} we see that few giant twins were found, but unfortunately AMBRE parameters for them were unavailable. 
\end{itemize}
{ We state here that our analysis based on EWs is not intended to derive atmospheric parameters. The quantities we look at, such as the slope of neutral lines EWs or the difference between ionised and neutral lines EWs, are related to the differences in the atmospheric parameters. We know that atmospheric parameters are crucial for distinguishing the different kind of stars from the spectra. Moreover, from e.g. regular spectroscopic parallax methods, we also know that these parameters are important  for distance determination.}

\end{appendix}

\end{document}